\documentclass[twocolumn,showpacs,preprintnumbers,prb,fleqn,floatfix]{revtex4}

\usepackage{graphicx}
\usepackage{dcolumn}
\usepackage{bm}
\usepackage{amsmath, amssymb}

\begin{document}

\title{Interplay among spin, orbital effects and localization in a GaAs two-dimensional electron gas in a strong in-plane magnetic field}

\author{B. A. Piot$^{1}$, D. K. Maude$^{1}$, U. Gennser $^{2}$, A. Cavanna $^{2}$, and D. Mailly $^{2}$}

\affiliation{$^{1}$ Laboratoire National des Champs Magn\'etiques
Intenses, Grenoble High Magnetic Field Laboratory, Centre National
de la Recherche Scientifique, 25 Avenue des Martyrs, F-38042
Grenoble, France}

\affiliation{$^{2}$ Laboratoire de Photonique et de
Nanostructures, Centre National de la Recherche Scientifique,
Route de Nozay, 91460 Marcoussis, France}

\date{\today }

\begin{abstract}

The magnetoresistance of a low carrier density, disordered GaAs
based two-dimensional (2D) electron gas has been measured in
parallel magnetic fields up to 32~T. The feature in the resistance
associated with the complete spin polarization of the carriers
shifts down by more than 20~T as the electron density is reduced,
consistent with recent theories taking into account the
enhancement of the electron-electron interactions at low
densities. Nevertheless, the magnetic field for complete
polarization, $B_{p}$, remains 2-3 times smaller than predicted
for a disorder free system. We show, in particular by studying the
temperature dependance of $B_{p}$ to probe the effective size of
the Fermi sea, that localization plays an important role in
determining the spin polarization of a 2D electron gas.
\end{abstract}
 \maketitle

\section {INTRODUCTION}

The subtle physics of 2D systems subjected to a parallel magnetic
field has stimulated increasing interest over the last decade.
Historically, the selective coupling of the in-plane magnetic
field to the electronic spin degree of freedom has been used to
probe the influence of many body effects on the spin polarization
and the ground state of the 2D electron gas (2-DEG). In practice,
the application of a parallel magnetic field progressively lifts
the spin degeneracy of the Fermi liquid until it becomes fully
spin polarized. In transport experiments for example, the
associated reduction in the screening of charged impurities leads
to an increase of the longitudinal resistance followed by a
saturation or a `kink' signaling the complete spin polarization of
the 2D system. \cite{Okamoto99,Vitkalov00,Tutucelectron02} The
magnetic field $B_{p}$ required to achieve complete polarization
is a subtle quantity depending on the magnetic field dependent
susceptibility, which is itself renormalized by many body effects
(see e.g. Ref.~[\onlinecite{zhangBp}]). An examination of the
dependence of $B_{p}$ on different physical parameters such as
carrier density, disorder, and temperature, has mostly been
carried out in Si and p-GaAS based 2D systems, leaving the
otherwise widely investigated n-GaAs 2DEG relatively unexplored
concerning these issues.

However, it was recently emphasized that in non-ideal 2D systems
which have a significant width, which is particularly relevant for
n-GaAs, the in-plane magnetic field also couples to the orbital
motion, and can considerably modify the physics involved, changing
the effective disorder of the system, \cite{DasSarmaHwang} even
stabilizing new phases. \cite{PIOTnatphys} It is therefore
important to characterize the various physical effects generated
by the presence of a parallel magnetic field, in order to gain a
better understanding of the influence of electron-electron
interactions on the ground state properties of the 2-DEG.

In this work, we report on experiments where a large in-plane
magnetic field is applied to a 2D \textit{electron} gas in GaAs,
which reveal the interplay between disorder, spin and, orbital
physics. Transport measurements under parallel magnetic fields up
to $32$~T show that, at sufficiently low electron density, the
signatures of orbital, spin effects and localization coexist. Such
a coexistence of distinct phenomena in a 2D system has to our
knowledge not been previously reported. The magnetoresistance
kink, associated with the complete spin polarization of the 2-DEG,
is observed for magnetic fields up to $B_{p} \sim 30$~T. The
observed reduction of $B_{p}$ of more than 20~T with decreasing
electron density (and mobility) is in good agreement with the
predicted enhancement of the spin susceptibility at low density
due to electron-electron interactions. However, the absolute value
of $B_{p}$ remains 2-3 times smaller than the calculated value at
$T=0K$ for a disorder-free system, and extrapolate to $B_{p}=0$ at
a large, non zero electron density. This behavior is attributed to
the localization of electrons in our disordered 2-DEG. $B_{p}$
shows a strong increase with temperature, consistent with a
reduction of the spin susceptibility. A simple model yields an
estimate of the density of delocalized states which is
significantly smaller than the total electron density, confirming
the role played by disorder in the observed phenomenon. In the
frozen spin regime, the resistivity $\rho$ dramatically increases
with a $ln(\rho)\propto B^{2}$ dependence, in agreement with
theory taking into account the orbital coupling to the parallel
field.

\section {EXPERIMENTAL DETAILS}

The sample studied here is a volume-doped
Al$_{x}$Ga$_{1-x}$As/GaAs heterojunction (HJ), patterned into a
Hall bar. The electron density $n_{s}$ was varied by the
application of a gate voltage enabling us to tune the density
between $4.2 \times 10^{10}cm^{-2}$ and $1.74 \times 10^{11}
cm^{-2}$ so that the corresponding interaction parameter $r_{s}$
spans the range $3 > r_{s} > 1.5$. At zero gate voltage the
mobility is $~38 m^{2}/V s$, and decreases down to $\sim 2 m^{2}/V
s$ in the low density regime (see Table~\ref{tab:table1}). The
transport measurements were performed with a standard low
frequency lock-in technique for temperatures between 1.17K and
4.2K, under magnetic fields up to 32 T produced by a resistive
magnet. The 2D electron plane was aligned parallel to the magnetic
field direction by using an in-situ rotation stage to minimize the
Hall voltage, with an accuracy estimated to be better than
$~1^{\circ}$. In the geometry used, the current along the Hall bar
was always perpendicular to the applied magnetic field. The
physical sample parameters are summarized for selected densities
in Table \ref{tab:table1}. Here the carrier densities are
determined from the gate voltage versus carrier density dependence
determined (for the same cooldown) in perpendicular magnetic field
from the Hall voltage and/or the Shubnikov de Haas effect.

\begin{table}[b]
\centering \caption{Parameters of the gated GaAs HJ .Electron density ($n_{s}$), zero field resistivity ($\rho_{0}$),
mobility ($\mu$) and magnetic field for full polarization ($B_{p}$) all measured at $T=1.17$~K.}\label{tab:table1}
\begin{tabular}{c|cccc}
 $n_{s}$~(cm$^{-2}$) & $\rho_{0}$ ($k\Omega/\square$)& $\mu$ ($m^{2}/Vs$) & $B_{p}$ (T) \\
\hline
 $5.69 \times 10^{10}$ & $2.43$ & 4.53 & 26.0 \\
 $5.14 \times 10^{10}$ & $3.86$ & 3.15 & 18.3 \\
 $4.76 \times 10^{10}$ & $5.70$ & 2.30 & 11.8 \\
 $4.31\times 10^{10}$ & $8.01$ & 1.81 & 6.2\\
\end{tabular}
\end{table}

\section {RESULTS AND DISCUSSION}

\subsection {Spin-determined magnetoresistance}

\begin{figure}[tbp]
\includegraphics[width=0.85\linewidth,angle=0,clip]{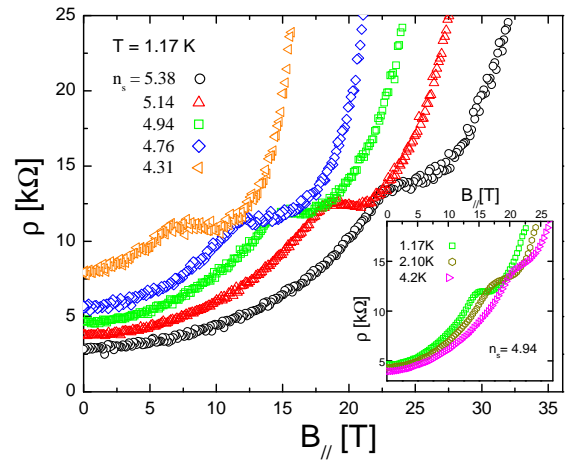}
\caption{(Color online) Longitudinale resistivity $\rho$ as a
function of the parallel magnetic field for different electron
densities $n_{s}$ in $10^{10} cm^{-2}$. $T=1.17K$. Inset:
temperature dependance of $\rho(B_{\parallel})$ at $n_{s}=4.94
\times 10^{10} cm^{-2}$.}\label{fig1}
\end{figure}

In Fig.\ref{fig1}, we plot the longitudinal resistivity $\rho$ as
a function of the parallel magnetic field for different electron
densities in the low density region. A number of different
magnetic field regimes are observed over the range studied
[0-32T]. The resistivity first increases as  $\rho \propto B^{2}$
(as observed in Refs.~[\onlinecite{Okamoto99,Vitkalov00}]) until a
kink emerges at a density-dependent magnetic field. This kink has
been shown to be associated with the complete spin polarization of
the system \cite{Tutucelectron02} eventually achieved at
sufficiently high parallel magnetic field. The rich physics lying
behind this quantity will be discussed throughout this paper. In
2-DEG's in Si-MOSFET's, in which a similar feature was first
reported, \cite{Okamoto99,Vitkalov00} a saturation of the
resistance is usually observed once full spin polarization occurs.
Here, the resistance does not saturate after the kink but smoothly
increases before rising exponentially, a behavior we attribute to
the non-negligible thickness of the 2-DEG in GaAs making orbital
coupling to the parallel magnetic field possible.
\cite{SarmaHwang05} We stress this behaviour, which will be
discussed in section \ref{SectionIIIB}, is qualitatively and
quantitatively different from the initial $\rho \propto B^{2}$
increase.

\begin{figure}[tbp]
\includegraphics[width=0.85\linewidth,angle=0,clip]{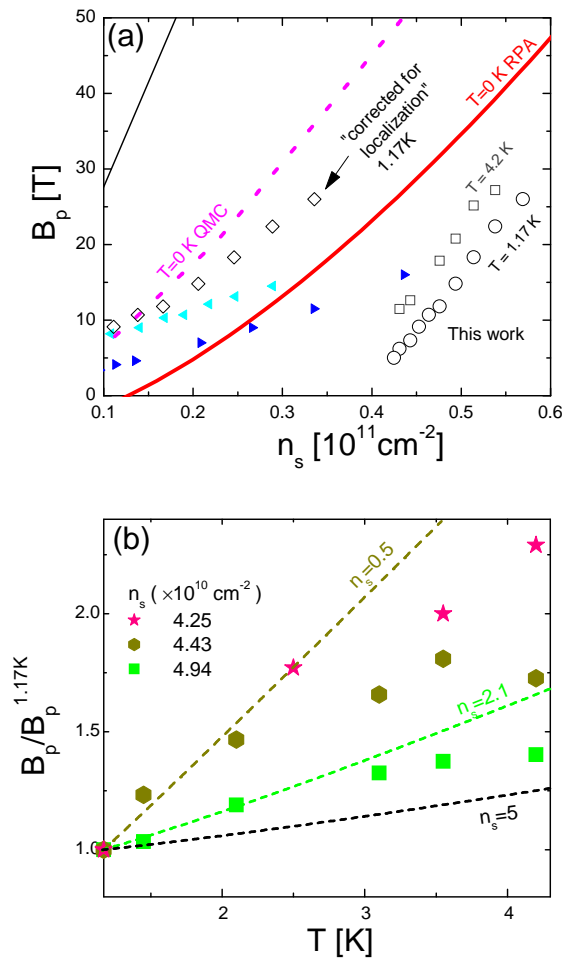}
\caption{(Color online) (a) Complete polarization field $B_{p}$ as
a function of the electron density. Data at $T=1.17K$ and $T=4.2K$
from this work (open circles and open squares, respectively), and
at $T=25mK$ from Refs.
~[\onlinecite{Tutucelectron02,Tutucwidth03}] (solid left and right
triangles, respectively.) Data at $T=1.17K$ plotted as a function
of the ``corrected'' electron density (diamonds) (see text).
Theoretical $T=0K$ calculations of the magnetic field for full
spin polarization in a single particle picture (solid line) and
including many-body effects: RPA \cite{zhangBp} (thick solid
line), and QMC \cite{TanatarBpara} (dotted line) (see text). (b)
Temperature dependence of $B_{p}$ normalized with respect to the
value at $1.17$~K, measured for different densities given in
$10^{10} cm^{-2}$ (symbols). Calculation in a single particle
picture for $n_{s}=0.5, 2.1 $ and $5 \times 10^{10} cm^{-2}$
(dotted line) (see text).}\label{fig2}
\end{figure}

From the magnetoresistance traces, we can extract the field
$B_{p}$ at which a kink occurs in the resistivity. In Fig.
\ref{fig2}(a), we plot the values of this field at $T=1.17K$ and
$T=4.2K$ as a function of the electron density, together with data
from Refs.~[\onlinecite{Tutucelectron02,Tutucwidth03}] obtained at
lower density in high mobility n-GaAs 2D systems. In a single
electron picture, for an ideal 2-DEG, the condition for the full
polarization of the Fermi sea is simply $g^{*}\mu_{B}B\geq2E_{F}$,
where $g^{*}$ is the electronic effective bare g-factor, $\mu_{B}$
the Bohr magneton and $E_{F}$ the Fermi energy. Due to the strong
enhancement of the g-factor by electron-electron exchange
interactions at low density, the magnetic field for full spin
polarization is actually much smaller. The calculation of this
quantity is a challenging task since it requires modelling
electron-electron interactions in the strongly interacting regime.
Furthermore, the polarization dependance of the exchange
interaction leads to a spin susceptibility which evolves non
linearly with the magnetic field, thus affecting the polarization
process.\cite{zhangBp} We plot in Fig.\ref{fig2}(a) the critical
magnetic field for full polarization at $T=0K$ in the single
electron picture (solid black line), together with recent
theoretical calculations including many-body effects using
different methods, respectively Random Phase Approximation
(RPA)\cite{zhangBp} and Quantum Monte Carlo Calculation (QMC).
\cite{TanatarBpara}

Our data at $1.17 K$ shows a dramatic reduction of $B_{p}$ as the
density is lowered, the latter decreasing by 21T in the range
$[4.25-5.69]\times 10^{10}cm^{-2}$. This reduction with decreasing
density is in good agreement with the $T=0K$ predictions of Ref.
~[\onlinecite{zhangBp}] , as well as the predictions of
Ref.~[\onlinecite{TanatarBpara}], where $B_{p}$ also drops by
about 20T over the same density range. However, there is a global
offset compared to theory, with the measured $B_{p}$ being about 2
or 3 times smaller than predicted. Furthermore, the low
temperature $B_{p}$ versus $(n_{s})$ trend extrapolates to zero at
a density of about $\sim 3.5 \times 10^{10}cm^{-2}$, apparently
suggesting the appearance of a ferromagnetic state, whereas no
ferromagnetic state has ever been observed in GaAs at these
densities in much cleaner samples, even at mK temperature (see
e.g. Ref. ~[\onlinecite{Zhu03}]). From a theoretical point of
view, this transition is also expected for much lower densities
(higher $r_{s}$ parameter) (see e.g. Refs.
~[\onlinecite{Attaccalite,Sarmaferro}]). We believe that claims of
a finite density ferromagnetic instability based only on the
extrapolation of the $B_{p}(n_{s})$ curves to zero magnetic field
are misleading.

We argue that the physics that can account for these observations
lies in the influence of disorder which has so far been neglected.
The influence of disorder on the spin polarization has already
been discussed for Si MOSFET's.
\cite{Vitkalovdisorder02,Pudalovdisorder02} In Ref.
~[\onlinecite{Pudalovdisorder02}], the field at which the
magnetoresistance saturates was found to be sample dependent,
increasing with the sample mobility. It was subsequently argued
that the magneto resistance saturation actually reflects complete
spin polarization of the \textit{non-localized} carriers only,
\cite{Goldcommentonpudalov} so that the value of $B_{p}$ is very
sensitive to the number of localised states in the system. At zero
perpendicular magnetic field, a ``2-component model'', where a
certain fraction of the electron are considered ``bound'' to
impurities (i.e. localized), was introduced in Ref.
~[\onlinecite{DasSarmaMIT}] in the context of the 2D Metal-to-
Insulator transition (MIT), giving good agreement with
experimental results. In our disordered GaAs 2-DEG, in which a
significant fraction of electrons are localized, the Fermi sea
which the magnetic field ``has to polarize'' is then effectively
smaller, resulting in an experimental $B_{p}$ smaller than the one
predicted by theories neglecting localization.

Additionally, as the density is reduced from $5.69\times
10^{10}cm^{-2}$ to $4.25\times 10^{10}cm^{-2}$ the electron
mobility in our sample drops by a factor of nearly 3 (see Table
\ref{tab:table1}). The associated decrease of the conductivity
$\sigma_0 = 1/\rho_{0}$ at T$=1.17$K as a function of the electron
density can be well described using the percolation model for the
apparent MIT (see e.g.\cite{He98percomit}), where $\sigma_0
\propto (n_{s}-n_{c})^{p}$, with $p=1.31$ the conductivity
exponent and $n_{c}$ the critical density for the MIT. Using
$n_{c}$ as the only adjustable parameter, one finds
$n_{c}=4.56\times 10^{10}cm^{-2}$, \footnote{It is important to
stress here that a precise location of the MIT cannot be inferred
from our T=1.17K data. The critical density $n_{c}$ we quote here
is only an approximation of the zero temperature value not taking
into account the temperature dependence of the conductivity
exponent, p.} which suggests that in this density region a
significant proportion of the electronic states become localized,
consistent with $B_{p}$ falling to zero in this region.

We believe these disorder effects are not observed in the data of
Refs. ~[\onlinecite{Tutucelectron02,Tutucwidth03}] because of the
much higher mobility (at least an order of magnitude) of the
sample investigated therein. In the latter work, the experimental
data and the theoretical predictions merge as density is lowered
(see Tutuc et al. data in Fig.\ref{fig2}(b)). It has been
suggested that this behavior can be understood by taking into
account the effective mass increase due to the orbital coupling to
the parallel field in a 2-DEG with finite thickness. The increased
effective mass leads to a smaller Fermi energy and thus a
reduction of $B_{p}$. In such a picture the deviation from the
theory for an ideal 2-DEG is more pronounced at high electron
density where the parallel magnetic fields involved are larger.
Such a behavior is not observed in our data, in which the
discrepancies between theory and experiment only weakly depend on
density and are actually slightly more pronounced at lower
densities. This confirms that in our case the $B_{p}$ reduction is
a disorder-induced phenomenon rather than a finite thickness
effect, the latter being most likely overwhelmed by the dramatic
effect of electron localization (weaker in Refs.
~[\onlinecite{Tutucelectron02,Tutucwidth03}]).

The effect of temperature on the magnetoresistance is shown for a
density of $4.94\times 10^{10}cm^{-2}$ in the inset of
Fig.\ref{fig1}, where a strong increase of $B_{p}$ is observed
between 1.17~K and 4.2~K. The evolution of $B_{p}$ with respect to
its value at $T=1.17$~K, is plotted as a function of temperature
in Fig.\ref{fig2}(b). As can be seen, in this temperature region,
$B_{p}$ follows a sub-linear variation with T, the initial fast
increase progressively slowing down at higher temperature. This
trend differs from the temperature dependence of $B_{p}$ in the
very low temperature regime recently observed for a 2D hole gas
.\cite{gao06} The initial enhancement of $B_{p}$ becomes more and
more pronounced as the density is decreased in the range
$[4.25-5.14]\times 10^{10} cm^{-2}$ (only 3 densities are shown in
Fig.\ref{fig2}(b) for clarity). The primary influence of
temperature at a given magnetic field is to reduce the electron
spin polarization. In a very simple single electron picture, the
spin polarization $S$ of the 2D Fermi sea in a parallel magnetic
field is given by

\begin{multline*}
S(n_{s},\Delta,E_{F},T) = \\ \frac{m^{*}}{2\pi\hbar^{2}n_{s}}  \int_{0}^{\infty} FD(E,E_{F},T)\Phi(E)  dE \\
- \frac{m^{*}}{2\pi\hbar^{2}n_{s}} \int_{\Delta}^{\infty}
FD(E,E_{F},T)\Phi(E-\Delta) dE,
\end{multline*}

where $FD$ is the Fermi distribution function, $\Phi$ is a
Heaviside step function, $\Delta$ the spin gap separating the spin
up and down subbands, $m^{*}$ the electron effective mass, and
$E_{F}$ the magnetic field and temperature-dependent Fermi energy.
One expects the magnetic field $B_{p}$, corresponding to $ S = 1$,
to increase with temperature to compensate for the polarization
reduction induced by the smearing of the Fermi distribution at the
Fermi level.

Calculations of $B_{p} (T)$ normalized with respect to its value
at $1.17$~K obtained using this formalism are shown by the dotted
lines in Fig.\ref{fig2}(b), for different electron densities. At
higher electron density, the Fermi sea is more robust with respect
to temperature so that the $B_{p}$ enhancement is less important,
in qualitative agreement with our experimental results.
\cite{highTdev} Intriguingly, the quantitative temperature
dependence corresponds more to the expected dependence for low
density systems ($n_{s} \sim 0.5-2.1 \times 10^{10} cm^{-2}$). We
take this as evidence that only a fraction of the total electron
density contributes to the effective ``delocalized'' Fermi sea
responsible for the thermal effects, reinforcing our
``disorder-based'' interpretation of the reduced $B_{p}$ values.
Comparing the experimental and theoretical low temperature
dependence, we can estimate the number of \textit{delocalized
}states in our system. For instance, for $n_{s}=4.94\times 10^{10}
cm^{-2}$ (solid squares in Fig.\ref{fig2}(b)), the electron
density that best reproduces the low temperature trend using our
model is $=2.1\times 10^{10} cm^{-2}$. This suggests that only
$~40\%$ of the electrons are delocalized at this density. For a
total density of $n_{s}=4.25\times 10^{10} cm^{-2}$ (hexagons in
Fig.\ref{fig2}(b)) our model yields a density of only $=0.5\times
10^{10} cm^{-2}$, meaning $~90\%$ of the states have become
localized, consistent with the proximity of the MIT in this
density region. \cite{exchange} We can use the estimated density
of \textit{delocalized} electrons as a function of $n_{s}$ and
re-plot the experimental $B_{p}$ at 1.17K as a function of these
values (diamonds in Fig.\ref{fig2}.a), which gives a better
quantitative agreement with theory. We stress that, unlike the
original data set (open circles), this corrected data set does not
extrapolate to $B_{p}$=0 for a non zero carrier density (not
shown). There is therefore no evidence of a ferromagnetic
transition once localization is taken into account.

\subsection {High magnetic field behavior}\label{SectionIIIB}

\begin{figure}[tbp]
\includegraphics[width=0.85\linewidth,angle=0,clip]{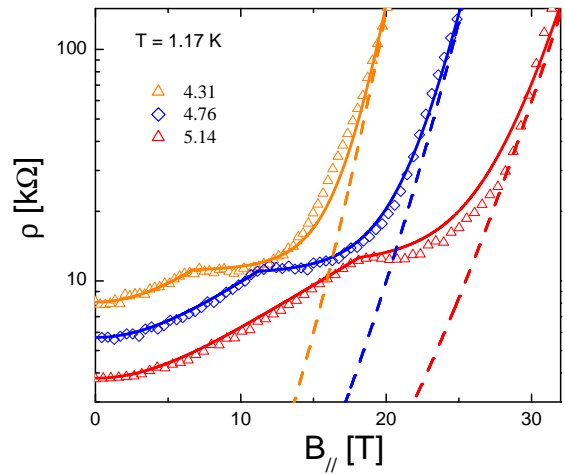}
\caption{(Color online) Resistivity $\rho$ in the high parallel
field regime at $T=1.17K$, for different densities $n_{s}$ in
$\times 10^{10} cm^{-2}$ (symbols). $ln(\rho)\propto B^{2}$
(dashed lines) and qualitative trend combining spin and orbital
effects (solid lines) (see text).}\label{fig3}
\end{figure}

We now turn to the description of the ``frozen-spin'' regime where
the system is fully polarized. As can be seen in Fig.\ref{fig1},
for $B > B_{p}$, the resistance increases slowly before rising
abruptly as the magnetic field is further increased. We stress
that in this case, as the spin degree of freedom is frozen, the
physical origin of this phenomenon cannot be spin related. In
Fig.\ref{fig3} we focus on the high field behavior of the
resistance by plotting the resistivity $\rho$ on a log scale. In
the high field limit, the resistivity asymptotically follows an
exponential power law, $ln(\rho)\propto \alpha (B)^{2}$, where
$\alpha$ increases monotonically as the density is reduced (dashed
lines in Fig.\ref{fig3}). The observed power law is qualitatively
consistent with an increased intersubband scattering due to the
orbital coupling to the parallel field, for which an identical
magnetic field dependance, $ln(\rho)\propto B^{2}$, is expected.
\cite{DasSarmaHwang} The orbital coupling to the in-plane magnetic
field can be thought of as a parallel field induced effective 2D
to 3D crossover in our low density quasi-2D system. The fact that
the coefficient $\alpha$ increases at low density is consistent
with this effect being stronger as the system becomes more dilute
.\cite{DasSarmaHwang} One can combine these orbital effects with
the initial $\rho \propto B^{2}$ ``spin trend'' to try to
reproduce the general behavior of the resistance. For this we use
the following expression for $\rho$, $\rho=\rho_{0}+ \beta B^{2} +
\gamma e^{\alpha B^{2}}$, where the second and third term
corresponds to spin and orbital contributions to the resistance
respectively. The ``spin term'' is set to $\beta {B_{p}}^{2}$ for
$B > B_{p}$ to account for the quench of spin effects, and
$\beta=0.024-0.065$ increases with decreasing density, consistent
with spin polarization growing faster at lower electron density.
The ``orbital term'' is determined from the high field limit, with
$\gamma=0.1$ (independent of density) and $\alpha=0.007-0.018$ for
our density range. The obtained behaviour is plotted as solid
lines in Fig.\ref{fig3} and reproduces the experimental data very
well leading further support to our simple model.

\section {CONCLUSION}

In conclusion, we have shown that the physics of GaAs 2D electron
systems under a strong parallel magnetic field is at low electron
density determined by an \textit{interplay} between spin,
localization and orbital effects. The magnetic field at which
complete spin polarization occurs is dramatically reduced both by
electron localization and electron-electron interaction enhanced
at low carrier density. The large value of the extrapolated
carrier density for full spin polarization at zero magnetic field
simply reflects the large number of localized electrons present in
disordered systems and should under no circumstances be
interpreted as evidence for a ferromagnetic instability.

\acknowledgements{The work at GHMFL was partially supported by the European 6th Framework Program under contract number
RITA-CT-3003-505474}


\begin{thebibliography}{19}
\expandafter\ifx\csname
natexlab\endcsname\relax\def\natexlab#1{#1}\fi
\expandafter\ifx\csname bibnamefont\endcsname\relax
  \def\bibnamefont#1{#1}\fi
\expandafter\ifx\csname bibfnamefont\endcsname\relax
  \def\bibfnamefont#1{#1}\fi
\expandafter\ifx\csname citenamefont\endcsname\relax
  \def\citenamefont#1{#1}\fi
\expandafter\ifx\csname url\endcsname\relax
  \def\url#1{\texttt{#1}}\fi
\expandafter\ifx\csname
urlprefix\endcsname\relax\def\urlprefix{URL }\fi
\providecommand{\bibinfo}[2]{#2}
\providecommand{\eprint}[2][]{\url{#2}}

\bibitem[{\citenamefont{Okamoto et~al.}(1999)\citenamefont{Okamoto, Hosoya,
  Kawaji, and Yagi}}]{Okamoto99}
\bibinfo{author}{\bibfnamefont{T.}~\bibnamefont{Okamoto}},
  \bibinfo{author}{\bibfnamefont{K.}~\bibnamefont{Hosoya}},
  \bibinfo{author}{\bibfnamefont{S.}~\bibnamefont{Kawaji}}, \bibnamefont{and}
  \bibinfo{author}{\bibfnamefont{A.}~\bibnamefont{Yagi}},
  \bibinfo{journal}{Phys. Rev. Lett.} \textbf{\bibinfo{volume}{82}},
  \bibinfo{pages}{3875} (\bibinfo{year}{1999}).

\bibitem[{\citenamefont{Vitkalov et~al.}(2000)\citenamefont{Vitkalov, Zheng,
  Mertes, Sarachik, and Klapwijk}}]{Vitkalov00}
\bibinfo{author}{\bibfnamefont{S.~A.} \bibnamefont{Vitkalov}},
  \bibinfo{author}{\bibfnamefont{H.}~\bibnamefont{Zheng}},
  \bibinfo{author}{\bibfnamefont{K.~M.} \bibnamefont{Mertes}},
  \bibinfo{author}{\bibfnamefont{M.~P.} \bibnamefont{Sarachik}},
  \bibnamefont{and} \bibinfo{author}{\bibfnamefont{T.~M.}
  \bibnamefont{Klapwijk}}, \bibinfo{journal}{Phys. Rev. Lett.}
  \textbf{\bibinfo{volume}{85}}, \bibinfo{pages}{2164} (\bibinfo{year}{2000}).

\bibitem[{\citenamefont{Tutuc et~al.}(2002)\citenamefont{Tutuc, Melinte, and
  Shayegan}}]{Tutucelectron02}
\bibinfo{author}{\bibfnamefont{E.}~\bibnamefont{Tutuc}},
  \bibinfo{author}{\bibfnamefont{S.}~\bibnamefont{Melinte}}, \bibnamefont{and}
  \bibinfo{author}{\bibfnamefont{M.}~\bibnamefont{Shayegan}},
  \bibinfo{journal}{Phys. Rev. Lett.} \textbf{\bibinfo{volume}{88}},
  \bibinfo{pages}{036805} (\bibinfo{year}{2002}).

\bibitem[{\citenamefont{Zhang and Sarma}(2006)}]{zhangBp}
\bibinfo{author}{\bibfnamefont{Y.}~\bibnamefont{Zhang}} \bibnamefont{and}
  \bibinfo{author}{\bibfnamefont{S.~D.} \bibnamefont{Sarma}},
  \bibinfo{journal}{Phys. Rev. Lett.} \textbf{\bibinfo{volume}{96}},
  \bibinfo{eid}{196602} (\bibinfo{year}{2006}).

\bibitem[{\citenamefont{{Das Sarma} and Hwang}(2000)}]{DasSarmaHwang}
\bibinfo{author}{\bibfnamefont{S.}~\bibnamefont{{Das Sarma}}} \bibnamefont{and}
  \bibinfo{author}{\bibfnamefont{E.~H.} \bibnamefont{Hwang}},
  \bibinfo{journal}{Phys. Rev. Lett.} \textbf{\bibinfo{volume}{84}},
  \bibinfo{pages}{5596} (\bibinfo{year}{2000}).

\bibitem[{\citenamefont{Piot et~al.}(2008)\citenamefont{Piot, Jiang, Dean,
  Engel, Gervais, Pfeiffer, and West}}]{PIOTnatphys}
\bibinfo{author}{\bibfnamefont{B.~A.} \bibnamefont{Piot}},
  \bibinfo{author}{\bibfnamefont{Z.}~\bibnamefont{Jiang}},
  \bibinfo{author}{\bibfnamefont{C.~R.} \bibnamefont{Dean}},
  \bibinfo{author}{\bibfnamefont{L.~W.} \bibnamefont{Engel}},
  \bibinfo{author}{\bibfnamefont{G.}~\bibnamefont{Gervais}},
  \bibinfo{author}{\bibfnamefont{L.~N.} \bibnamefont{Pfeiffer}},
  \bibnamefont{and} \bibinfo{author}{\bibfnamefont{K.~W.} \bibnamefont{West}},
  \bibinfo{journal}{Nat Phys} \textbf{\bibinfo{volume}{4}},
  \bibinfo{pages}{936} (\bibinfo{year}{2008}).

\bibitem[{\citenamefont{Das~Sarma and Hwang}(2005)}]{SarmaHwang05}
\bibinfo{author}{\bibfnamefont{S.}~\bibnamefont{Das~Sarma}} \bibnamefont{and}
  \bibinfo{author}{\bibfnamefont{E.~H.} \bibnamefont{Hwang}},
  \bibinfo{journal}{Phys. Rev. B} \textbf{\bibinfo{volume}{72}},
  \bibinfo{pages}{035311} (\bibinfo{year}{2005}).

\bibitem[{\citenamefont{Tutuc et~al.}(2003)\citenamefont{Tutuc, Melinte, {E. P.
  De Poortere}, Shayegan, and Winkler}}]{Tutucwidth03}
\bibinfo{author}{\bibfnamefont{E.}~\bibnamefont{Tutuc}},
  \bibinfo{author}{\bibfnamefont{S.}~\bibnamefont{Melinte}},
  \bibinfo{author}{\bibnamefont{{E. P. De Poortere}}},
  \bibinfo{author}{\bibfnamefont{M.}~\bibnamefont{Shayegan}}, \bibnamefont{and}
  \bibinfo{author}{\bibfnamefont{R.}~\bibnamefont{Winkler}},
  \bibinfo{journal}{Phys. Rev. B.} \textbf{\bibinfo{volume}{67}},
  \bibinfo{pages}{241309(R)} (\bibinfo{year}{2003}).

\bibitem[{\citenamefont{Suba\c{s}i and Tanatar}(2008)}]{TanatarBpara}
\bibinfo{author}{\bibfnamefont{A.~L.} \bibnamefont{Suba\c{s}i}}
  \bibnamefont{and} \bibinfo{author}{\bibfnamefont{B.}~\bibnamefont{Tanatar}},
  \bibinfo{journal}{Phys. Rev. B} \textbf{\bibinfo{volume}{78}},
  \bibinfo{eid}{155304} (\bibinfo{year}{2008}).

\bibitem[{\citenamefont{Zhu et~al.}(2003)\citenamefont{Zhu, Stormer, Pfeiffer,
  Baldwin, and West}}]{Zhu03}
\bibinfo{author}{\bibfnamefont{J.}~\bibnamefont{Zhu}},
  \bibinfo{author}{\bibfnamefont{H.~L.} \bibnamefont{Stormer}},
  \bibinfo{author}{\bibfnamefont{L.~N.} \bibnamefont{Pfeiffer}},
  \bibinfo{author}{\bibfnamefont{K.~W.} \bibnamefont{Baldwin}},
  \bibnamefont{and} \bibinfo{author}{\bibfnamefont{K.~W.} \bibnamefont{West}},
  \bibinfo{journal}{Phys. Rev. Lett.} \textbf{\bibinfo{volume}{90}},
  \bibinfo{pages}{056805} (\bibinfo{year}{2003}).

\bibitem[{\citenamefont{Attaccalite et~al.}(2002)\citenamefont{Attaccalite,
  Moroni, Gori-Giorgi, and Bachelet}}]{Attaccalite}
\bibinfo{author}{\bibfnamefont{C.}~\bibnamefont{Attaccalite}},
  \bibinfo{author}{\bibfnamefont{S.}~\bibnamefont{Moroni}},
  \bibinfo{author}{\bibfnamefont{P.}~\bibnamefont{Gori-Giorgi}},
  \bibnamefont{and} \bibinfo{author}{\bibfnamefont{G.~B.}
  \bibnamefont{Bachelet}}, \bibinfo{journal}{Phys. Rev. Lett.}
  \textbf{\bibinfo{volume}{88}}, \bibinfo{pages}{256601}
  (\bibinfo{year}{2002}).

\bibitem[{\citenamefont{Zhang and Das~Sarma}(2005)}]{Sarmaferro}
\bibinfo{author}{\bibfnamefont{Y.}~\bibnamefont{Zhang}} \bibnamefont{and}
  \bibinfo{author}{\bibfnamefont{S.}~\bibnamefont{Das~Sarma}},
  \bibinfo{journal}{Phys. Rev. B} \textbf{\bibinfo{volume}{72}},
  \bibinfo{pages}{115317} (\bibinfo{year}{2005}).

\bibitem[{\citenamefont{Vitkalov et~al.}(2002)\citenamefont{Vitkalov, Sarachik,
  and Klapwijk}}]{Vitkalovdisorder02}
\bibinfo{author}{\bibfnamefont{S.~A.} \bibnamefont{Vitkalov}},
  \bibinfo{author}{\bibfnamefont{M.~P.} \bibnamefont{Sarachik}},
  \bibnamefont{and} \bibinfo{author}{\bibfnamefont{T.~M.}
  \bibnamefont{Klapwijk}}, \bibinfo{journal}{Phys. Rev. B}
  \textbf{\bibinfo{volume}{65}}, \bibinfo{pages}{201106}
  (\bibinfo{year}{2002}).

\bibitem[{\citenamefont{Pudalov et~al.}(2002)\citenamefont{Pudalov, Brunthaler,
  Prinz, and Bauer}}]{Pudalovdisorder02}
\bibinfo{author}{\bibfnamefont{V.~M.} \bibnamefont{Pudalov}},
  \bibinfo{author}{\bibfnamefont{G.}~\bibnamefont{Brunthaler}},
  \bibinfo{author}{\bibfnamefont{A.}~\bibnamefont{Prinz}}, \bibnamefont{and}
  \bibinfo{author}{\bibfnamefont{G.}~\bibnamefont{Bauer}},
  \bibinfo{journal}{Phys. Rev. Lett.} \textbf{\bibinfo{volume}{88}},
  \bibinfo{pages}{076401} (\bibinfo{year}{2002}).

\bibitem[{\citenamefont{Dolgopolov and Gold}(2002)}]{Goldcommentonpudalov}
\bibinfo{author}{\bibfnamefont{V.~T.} \bibnamefont{Dolgopolov}}
  \bibnamefont{and} \bibinfo{author}{\bibfnamefont{A.}~\bibnamefont{Gold}},
  \bibinfo{journal}{Phys. Rev. Lett.} \textbf{\bibinfo{volume}{89}},
  \bibinfo{pages}{129701} (\bibinfo{year}{2002}).

\bibitem[{\citenamefont{Das~Sarma and Hwang}(1999)}]{DasSarmaMIT}
\bibinfo{author}{\bibfnamefont{S.}~\bibnamefont{Das~Sarma}} \bibnamefont{and}
  \bibinfo{author}{\bibfnamefont{E.~H.} \bibnamefont{Hwang}},
  \bibinfo{journal}{Phys. Rev. Lett.} \textbf{\bibinfo{volume}{83}},
  \bibinfo{pages}{164} (\bibinfo{year}{1999}).

\bibitem[{\citenamefont{He and Xie}(1998)}]{He98percomit}
\bibinfo{author}{\bibfnamefont{S.}~\bibnamefont{He}} \bibnamefont{and}
  \bibinfo{author}{\bibfnamefont{X.~C.} \bibnamefont{Xie}},
  \bibinfo{journal}{Phys. Rev. Lett.} \textbf{\bibinfo{volume}{80}},
  \bibinfo{pages}{3324} (\bibinfo{year}{1998}).

\bibitem[{\citenamefont{Gao et~al.}(2006)\citenamefont{Gao, Boebinger,
  A.~P.~Mills, Ramirez, Pfeiffer, and West}}]{gao06}
\bibinfo{author}{\bibfnamefont{X.~P.~A.} \bibnamefont{Gao}},
  \bibinfo{author}{\bibfnamefont{G.~S.} \bibnamefont{Boebinger}},
  \bibinfo{author}{\bibfnamefont{J.}~\bibnamefont{A.~P.~Mills}},
  \bibinfo{author}{\bibfnamefont{A.~P.} \bibnamefont{Ramirez}},
  \bibinfo{author}{\bibfnamefont{L.~N.} \bibnamefont{Pfeiffer}},
  \bibnamefont{and} \bibinfo{author}{\bibfnamefont{K.~W.} \bibnamefont{West}},
  \bibinfo{journal}{Phys. Rev. B} \textbf{\bibinfo{volume}{73}},
  \bibinfo{eid}{241315} (\bibinfo{year}{2006}).

\bibitem[{hig()}]{highTdev}
\bibinfo{note}{We note that a sub-linear $B_{p}(T)$ trend is however not
  expected in our simple model, suggesting some more complexe effects are
  taking place at higher temperature}.

\bibitem[{exc()}]{exchange}
\bibinfo{note}{The ability of our single particle model to determine the actual
fraction of localized states is limited by the presence of many
body effects. Simple considerations based on self-consistent
exchange enhancement of the spin gap show that neglecting these
effects leads to an overestimation of the fraction of localized
states.}



\end{thebibliography}
\end{document}